\input epsf
\newfam\scrfam
\batchmode\font\tenscr=rsfs10 \errorstopmode
\ifx\tenscr\nullfont
	\message{rsfs script font not available. Replacing with calligraphic.}
\else	\font\sevenscr=rsfs7 
	\font\fivescr=rsfs5 
	\skewchar\tenscr='177 \skewchar\sevenscr='177 \skewchar\fivescr='177
	\textfont\scrfam=\tenscr \scriptfont\scrfam=\sevenscr
	\scriptscriptfont\scrfam=\fivescr
	\def\scr{\fam\scrfam}
	\def\cal{\scr}
\fi
\newfam\msbfam
\batchmode\font\twelvemsb=msbm10 scaled\magstep1 \errorstopmode
\ifx\twelvemsb\nullfont\def\Bbb{\bf}
	\message{Blackboard bold not available. Replacing with boldface.}
\else	\catcode`\@=11
	\font\tenmsb=msbm10 \font\sevenmsb=msbm7 \font\fivemsb=msbm5
	\textfont\msbfam=\tenmsb
	\scriptfont\msbfam=\sevenmsb \scriptscriptfont\msbfam=\fivemsb
	\def\Bbb{\relax\expandafter\Bbb@}
	\def\Bbb@#1{{\Bbb@@{#1}}}
	\def\Bbb@@#1{\fam\msbfam\relax#1}
	\catcode`\@=\active
\fi
\font\eightrm=cmr8		\def\xrm{\eightrm}
\font\eightbf=cmbx8		\def\xbf{\eightbf}
\font\eightit=cmti8		\def\xit{\eightit}
\font\eighttt=cmtt8		\def\xtt{\eighttt}
\font\eightcp=cmcsc8
\font\eighti=cmmi8		\def\xold{\eighti}
\font\teni=cmmi10		\def\old{\teni}

\font\tentt=cmtt10
\font\twelverm=cmr12
\font\twelvecp=cmcsc10 scaled\magstep1
\font\fourteencp=cmcsc10 scaled\magstep2

\def\ss{\scriptstyle}

\headline={\ifnum\pageno=1\hfill\else
{\eightcp Martin Cederwall: 
	``Boundaries of 11-Dimensional Membranes''}
		\dotfill{ }{\old\folio}\fi}
\def\makeheadline{\vbox to 0pt{\vss\noindent\the\headline\break
\hbox to\hsize{\hfill}}
	\vskip2\baselineskip}
\def\makefootline{\ifnum\foottest=1
	\baselineskip=.8cm\line{\the\footline}\global\foottest=0
	\fi
        }
\newcount\foottest
\foottest=0
\def\footnote#1#2{${\,}^#1$\footline={\vtop{\baselineskip=9pt
        \hrule width.5\hsize\hfill\break
        \indent ${}^#1$ \vtop{\hsize=14cm\noindent\xrm #2}}}\foottest=1
        }
\newcount\refcount
\refcount=0
\newwrite\refwrite
\def\ref#1#2{\global\advance\refcount by 1
	\xdef#1{{\old\the\refcount}}
	\ifnum\the\refcount=1
	\immediate\openout\refwrite=\jobname.refs
	\fi
	\immediate\write\refwrite
		{\item{[{\xold\the\refcount}]} #2\hfill\par\vskip-2pt}}
\def\refout{\catcode`\@=11 
	\xrm\immediate\closeout\refwrite
	\vskip2\baselineskip
	{\noindent\twelvecp References}\hfill\vskip\baselineskip
	\baselineskip=.75\baselineskip
	\input\jobname.refs 
	\baselineskip=4\baselineskip \divide\baselineskip by 3
	\catcode`\@=\active\rm}
\newcount\eqcount
\eqcount=0
\def\Eqn#1{\global\advance\eqcount by 1
	\xdef#1{
			{\old\the\eqcount}}
		\eqno(
			{\oldstyle\the\eqcount})}
\def\eqn{\global\advance\eqcount by 1
	\eqno(
		{\oldstyle\the\eqcount})}
\def\multi{\global\advance\eqcount by 1}
\def\multieq#1#2{\xdef#1{{\old\the\eqcount#2}}
	\eqno{({\oldstyle\the\eqcount#2})}}
\parskip=3.5pt plus .3pt minus .3pt
\baselineskip=14pt plus .1pt minus .05pt
\lineskip=.5pt plus .05pt minus .05pt
\lineskiplimit=.5pt
\abovedisplayskip=18pt plus 2pt minus 2pt
\belowdisplayskip=\abovedisplayskip
\hsize=15cm
\vsize=19cm
\hoffset=1cm
\voffset=1.8cm
\def\/{\over}
\def\*{\partial}
\def\a{\alpha}
\def\b{\beta}
\def\c{\gamma}
\def\d{\delta}
\def\e{\varepsilon}
\def\g{\gamma}
\def\k{\kappa}
\def\l{\lambda}
\def\x{\xi}
\def\w{\omega}
\def\D{\Delta}
\def\G{\Gamma}
\def\LL{{\cal L}}
\def\L{\Lambda}
\def\M{{\cal M}}
\def\dM{{\*\!\M}}
\def\Z{{\Bbb Z}}
\def\punkt{\,\,.}
\def\komma{\,\,,}
\def\minus{\!-\!}
\def\+{\!+\!}
\def\={\!=\!}
\def\half{{\lower2.5pt\hbox{\eightrm 1}\/\raise2.5pt\hbox{\eightrm 2}}}
\def\fraction#1{{\lower2.5pt\hbox{\eightrm 1}\/\raise2.5pt\hbox{\eightrm #1}}}
\def\Fraction#1#2{{\lower2.5pt\hbox{\eightrm #1}\/
	\raise2.5pt\hbox{\eightrm #2}}}
\def\tr{\hbox{\rm tr}}
\def\eg{{\tenit e.g.}}

\def\SYM{super-Yang--Mills}
\def\WZ{Wess--Zumino}
\def\CS{\Omega_{\rm CS}}
\def\WZW{Wess--Zumino--Witten}

%
%
%
%

\null\vskip-1cm
\hbox to\hsize{\hfill G\"oteborg-ITP-97-9}
\hbox to\hsize{\hfill\tt hep-th/9704161}
\hbox to\hsize{\hfill April, 1997}

\vskip1cm
\centerline{\epsffile{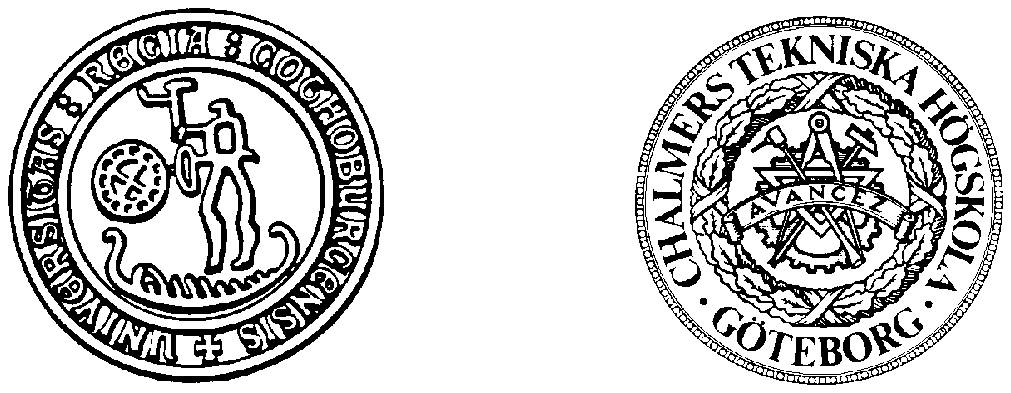}}

\vskip2cm
\centerline{\fourteencp Boundaries of 11-Dimensional Membranes} 
\vskip4pt
\vskip\parskip
\centerline{\twelvecp}

\vskip1.2cm
\centerline{\twelverm Martin Cederwall} 

\vskip.8cm
\centerline{\it Institute for Theoretical Physics}
\centerline{\it G\"oteborg University and Chalmers University of Technology }
\centerline{\it S-412 96 G\"oteborg, Sweden}

\vskip.8cm
\catcode`\@=11
\centerline{\tentt tfemc@fy.chalmers.se}
\catcode`\@=\active

\vskip2.2cm

\centerline{\bf Abstract}

{\narrower\noindent The action for an 11-dimensional supermembrane
contains a chiral Wess--Zumino--Witten model coupling to the 
E${}_8$ super-Yang--Mills theory on the end-of-the-world 9-brane.
It is demonstrated that this boundary string theory is dictated
both by gauge invariance and by $\k$-symmetry.\smallskip} 

\vfill
\eject

\def\nl{\hfill\break\indent}
\def\nlni{\hfill\break}

\ref\Horava{P. Ho\v rava and E. Witten, 
	\xit ``Heterotic and Type I String Dynamics from Eleven Dimensions'',
	\nl\xrm Nucl.Phys. {\xbf B460} ({\xold1996}) {\xold506} 
		({\xtt hep-th/9603142});\nl
	\xit ``Eleven-Dimensional Supergravity on a Manifold with Boundary'',
	\nl\xrm Nucl.Phys. {\xbf B475} ({\xold1996}) {\xold94} 
		({\xtt hep-th/9510209}).}
\ref\Mtheory{E.~Witten, 
	{\xit ``String Theory Dynamics in Various Dimensions''},
	Nucl.~Phys.~{\xbf B443}
		({\xold1995}) {\xold85} ({\xtt hep-th/9503124});\nlni
	J.H.~Schwarz, {\xit ``The Power of M Theory''},
	Phys.~Lett.~{\xbf B367} ({\xold1996}) {\xold97}
	({\xtt hep-th/9510086});\nl
	{\xit ``Lectures on Superstring and M Theory Dualities''},
	{\xtt hep-th/9607201};\nlni
	A.~Sen, {\xit ``Unification of String Dualities''},
	{\xtt hep-th/9609176};\nlni
	E.~Sezgin, {\xit ``The M Algebra''}, 
	Phys.Lett. {\xbf B392} ({\xold1997}) {\xold323} 
		({\xtt hep-th/9609086});\nlni
	T.~Banks, W.~Fischler, S.H.~Shenker and L.~Susskind,
	\nl{\xit ``M Theory as a Matrix Model: a Conjecture''},
	Phys.Rev. {\xbf D55} ({\xold1997}) {\xold5112} 
		({\xtt hep-th/9610043}).}
\ref\Elevensg{E. Cremmer, B. Julia and  J. Scherk,
	{\xit ``Supergravity Theory in Eleven-Dimensions''},
	\nl Phys.Lett. {\xbf76B} ({\xold1978}) {\xold409};\nlni
	L.~Brink and P.~Howe, {\xit ``Eleven-Dimensional Supergravity 
	on the Mass-Shell in Superspace''},
	\nl Phys.~Lett.~{\xbf 91B} ({\xold1980}) {\xold384};\nlni
	E.~Cremmer and S.~Ferrara, 
	{\xit ``Formulation of Eleven-Dimensional Supergravity 
	in Superspace''},\nl Phys.~Lett.~{\xbf 91B} ({\xold1980}) {\xold61}.}
\ref\HBianchi{B.E.W. Nilsson and A. Tollsten,
{\xit ``Superspace Formulation of the Ten-Dimensional Coupled\nl
	Einstein--Yang--Mills System''},
	Phys.Lett.{\xbf181B} ({\xold1986}) {\xold63}.}
\ref\Alwis{S.P. de Alwis, {\xit ``A Note on Brane Tension and M-Theory''},
	Phys.~Lett.~{\xbf B388} ({\xold1996}) {\xold291}
	({\xtt hep-th/9607011}).}
\ref\Supermembrane{E.~Bergshoeff, E.~Sezgin and P.K.~Townsend, 
	\nl {\xit ``Supermembranes and Eleven-Dimensional Supergravity''},
	Phys.~Lett.~{\xbf B189} ({\xold1987}) {\xold75};
	\nl {\xit ``Properties of the Eleven-Dimensional Supermembrane 
	Theory''}, Ann.~Phys. {\xbf 185} ({\xold1988}) {\xold330}.}
\ref\WessZuminoWitten{E. Witten, {\xit ``Non-Abelian Bosonization in 
	Two-Dimensions''}, 
		Commun.Math.Phys. {\xbf92} ({\xold1984}) {\xold455}.}
\ref\WittenSD{E. Witten, {\xit ``Five-Brane Effective Action In M-Theory''},
	{\xtt hep-th/9610234}.}
\ref\Fivebranes{R.~G\"uven, 
	{\xit ``Black p-Brane Solutions of D=11 Supergravity Theory''}, 
	Phys.~Lett.~{\xbf B276} ({\xold1992}) {\xold49};\nlni
	E. Witten, {\xit ``Five-Brane Effective Action in M-Theory''},
	{\xtt hep-th/9610234};\nlni
	I. Bandos, K. Lechner, A. Nurmagambetov, P. Pasti, D. Sorokin and 
	M. Tonin,\nl 
	{\xit ``Covariant Action for the Super-Five-Brane of M-Theory''},
	{\xtt hep-th/9701149};\nlni
	M. Aganagic, J. Park, C. Popescu and J.H. Schwarz,\nl
	{\xit ``World-Volume Action of the M Theory Five-Brane''},
	{\xtt hep-th/9701166};\nlni
	P.S. Howe, E. Sezgin and P. C. West,\nl
	{\xit ``Covariant Field Equations of the M Theory Five-Brane''},
	{\xtt hep-th/9702008}.}

\frenchspacing

Ho\v rava and Witten [\Horava] have described the heterotic and type I string
theories as a compactification of M-theory [\Mtheory] on 
$S^1\!/\Z_2$ (an interval). The cancellation of anomalies 
in the low-energy 11-dimensional supergravity on this space relies
on the introduction of an extra E${}_8$ \SYM\ multiplet on each
component of the boundary.

At each boundary 9-brane, only the field components survive that are
even under reflections in the boundary. This projects the 11-dimensional
supergravity [\Elevensg] down to type I supergravity, to which the \SYM\ fields
on the boundary couple. The 3-form field strength
$h$, which is what survives of the 4-form $H$ on the boundary,
picks up an anomalous term in its Bianchi identity, $dh\!\sim\!\tr F^2$,
as is necessary in the coupling of \SYM\ to type I supergravity [\HBianchi].
This in turn implies boundary conditions on the $H$ field,
$$
H|_\dM=-{\k^2\/2\l^2}F^2	\Eqn\boundarycondition
$$
(the normalization of $H$ adopted here differs from ref. [\Horava] by
a factor $\sqrt2$, and is the natural one from a superspace point of view).
It should be pointed out, that eq. (\boundarycondition), which so far
has been seen as a boundary condition for the bosonic 4-form, now is 
promoted to a superspace identity, with the former as its leading
component.
Any transformation involving the \SYM\ fields induces a transformation
of the $H$ field. In ref. [\Horava], this is used to verify anomaly
cancellation in the entire model, including both the bulk and boundary
fields (although some questions remain about the gravitational part). 
As a consequence, the gravity and gauge couplings must be related
as $\l^3\=(2\pi)^{5/2}2\k^2$. They are both related to the string tension
$(2\pi\a')^{-1}$ as $2\k^2\=(2\pi)^8{\a'}^{9/2}$ and 
$\l^2\=(2\pi)^7{\a'}^3$ [\Alwis].

In this note, the analogous mechanism for the supermembrane propagating
in the 11-dimen\-sional supergravity background will be studied,
in the case where the membrane ends on the 10-dimensional boundary.

The ordinary action for the supermembrane [\Supermembrane] is
$$
I_0=-T\!\int_\M\! d^3\x\sqrt{-g}+T\!\int_\M\! C\punkt	\Eqn\oldaction
$$
where the membrane tension is $T\=(2\pi)^{-2}{\a'}^{-3/2}$,
and where all world-volume fields are constructed as pullbacks of
superfields in 11-dimensional supergravity.
The action (\oldaction) is invariant under $\k$-symmetry, as long as there
are no boundaries present, thus ensuring that the number of physical fermions
on the world-volume equals the number of transverse oscillations.
When there is a boundary $\dM$, there must
be additional terms to cancel both the gauge anomaly induced by
(\boundarycondition) and the $\k$-symmetry anomaly.
Both contributions originate from the \WZ\ term in the action.

After using the relations among the coupling constants, one finds that
the potential $C$ close to the boundary behaves as 
a Chern--Simons form (modulo total derivatives):
$$
C|_\dM=-{1\/8\pi T}\,\CS(A)=-{1\/8\pi T}\,\tr\,(dA+\Fraction23A^3)
\punkt\eqn
$$
The gauge anomaly of the membrane action is therefore
$$
\d_\L I_0=-{1\/8\pi}\int_\dM \!\tr\,\L dA	\punkt\eqn
$$
The $\k$-symmetry, on the other hand, transforms $C$ as 
$\d'_\k C\=\LL_\k C\=(i_\k d\+di_\k)C$
(the prime is only to distinguish gauge and $\k$-transformations), 
and the presence of a boundary generates the anomaly
$$
\d'_\k I_0=-{1\/8\pi}\int_\dM\! i_\k C= -{1\/8\pi}\int_\dM\! i_\k\CS(A)
	\punkt\eqn
$$
It is convenient to consider instead the modified $\k$-transformation
$\D_\k\=\d'_\k\minus\d_{i_\k A}$, under which the anomaly becomes
$$
\D_\k I_0=-{1\/8\pi}\int_\dM\!\tr\,i_\k FA		\punkt\Eqn\kappavar
$$
We will demonstrate that both these anomalies are cancelled
by contributions from a chiral level one \WZW\ model living on the boundary
string of the membrane. 

The coupling of a \WZW\ model [\WessZuminoWitten] to external gauge fields can
be accomplished by replacing the left-invariant Maurer--Cartan 1-form 
$\w\=g^{-1}dg$ by $g^{-1}Dg\=\w\minus A$, $D$ being the covariant derivative
for the right action of the gauge group. Gauge invariance under
$$
\eqalign{&\d_\L A=D\L\komma\cr
	 &\d_\L g=g\L\komma\cr}\eqn
$$
also requires the inclusion of a \WZ\ term and a Chern--Simons term 
for $A$, and the invariant action reads
$$
\eqalign{
I_A&={1\/8\pi}\int_\dM\!\tr \Bigl(\half(\w-A){*}(\w-A)-\w A\Bigr)
	+{1\/8\pi}\int_\M\!\Bigl(\CS(\w)-\CS(A)\Bigr)\cr
   &\equiv I_1-{1\/8\pi}\int_\M\!\CS(A)	\punkt}\Eqn\Aaction
$$
The last term is of course impossible in the present context,
where $A$ lives only on the boundary, but it plays exactly the same 
r\^ole as the \WZ\ term in the
membrane action, so by leaving it out we have verified gauge invariance
of the full membrane action $I\=I_0\+I_1$:
$$
I=-T\!\int_\M\! d^3\x\sqrt{-g}+\int_\M\! \Bigl(TC+\Fraction{1}{8$\ss\pi$}
		\CS(\w)\Bigr)
	+{1\/8\pi}\int_\dM\!\tr \Bigl(\half(\w-A){*}(\w-A)-\w A\Bigr)
	\punkt\eqn
$$
It may be in order to comment on chirality of the \WZW\ model.
The ordinary \WZW\ action, obtained from (\Aaction) by setting $A$ to zero,
leads to the equations of motion for $g$:
$\*_-j_+\=0\=\*_+j_-$, where $j_+\=\*_+gg^{-1}$, $j_-\=g^{-1}\*_-g$,
and where a 1-form $\b$ is decomposed in selfdual and anti-selfdual components
as $\b_\pm d\x^\pm\={1\/2}(\b\pm{*}\b)$. The two equations are equivalent.
The solution is 
$g(\x^+,\x^-)\=g_L(\x^+)g_R(\x^-)$, and it may consistently be truncated
by setting (\eg) $j_+\=0$.
When the $A$ field is present, the two equivalent versions of the
equations of motion for $g$ become
$\*_-J_+\=0\=D_+J_-+F_{+-}$, where $J_+\=D_+gg^{-1}$, $J_-\=g^{-1}D_-g$.
Now, the only possible chirality constraint is $J_+\=0$, since $J_-$
couples to the background fields. This is the model under consideration
here. The mechanism is completely analogous to the one at work for free
chiral bosons or selfdual forms in $4n\+2$ dimensions [\WittenSD].
The chirality constraint is invariant under all symmetries considered.

The $\k$-transformation of $A$ is given as $\d'_\k A\=\LL_\k A$, $A$ being
the pullback of the background superfield. The transformation of
$g$ has to be guessed, but there are not many possibilities. If
$\d'_\k g\=gi_\k A$, the modified $\k$-transformations act as
$$
\eqalign{
	&\D_\k A=i_\k F\komma\cr
	&\D_\k g=0\cr}
\eqn
$$
(since $g$ is defined in the bulk and $A$ only on the boundary, 
it is actually necessary to use the modified transformation $\D_\k$
for the $\k$-transformations to be well defined).
The transformation of the new terms in the membrane action is then
$$
\D_\k I_1={1\/8\pi}\int_\dM\!\tr\Bigl((i_\k F+*i_\k F)\w-*i_\k FA\Bigr)
	\komma\eqn
$$
which obviously cancels against eq. (\kappavar) if $*i_\k F\=-i_\k F$.
This is the last condition needed, and it follows from the equations
of motion for the background superfields (these are of course expected
to be needed for $\k$-symmetry) as follows.

In the 10-dimensional type I superspace associated with the boundary,
$F$ is a superspace 2-form. There is not any known off-shell superspace
formulation of 10-dimensional \SYM, but by imposing a set of constraints
on $F$, the field content is reduced down to the correct one and the 
theory is put on-shell. With $a,b,\ldots$ and $\a,\b,\ldots$ denoting
inertial frame vector and spinor indices, respectively, the relevant
constraint is $F_{\a\b}\=0$. Once it is imposed, the Bianchi identity
$(DF)_{\a\b\c}\!\sim\!(\g_a)_{(\a\b}{F^a}_{\c)}\=0$ may be solved as
$F_{a\a}\={(\g_a)_\a}^\b \psi_\b$. The parameter $\k$ is constrained
to obey 
$$
\k=\fraction2(1+\G)\k\komma\quad\G=-{1\/2\sqrt{-g}}\e^{ij}\g_{ij}\komma\eqn
$$
on the boundary, and this implies that 
$$
(i_\k F)_i=(\bar\psi\g_i\k)
=-{1\/\sqrt{-g}}g_{ij}\e^{jk}(\bar\psi\g_k\k)=-(*i_\k F)_i\komma\eqn
$$
which completes the proof of $\k$-symmetry.

We have found that the incorporation of a chiral level 1 E${}_8$ \WZW\ model 
on the boundary $\dM$ of the world-volume $\M$ of an 11-dimensional
supermembrane ending on an end-of-the-world 9-brane of M-theory
cancels the apparent gauge and $\k$-symmetry anomalies of the bulk
theory of the membrane. The cancellations rely on the same relation
between coupling constants that was used for the anomaly cancellation
in the background theory. The boundary conformal field theory is the
membrane analog of string endpoint Chan--Paton factors. 
As in the context of the heterotic string,
there are of course other possible formulations
of the boundary theory; the chiral current may \eg\ be fermionized.
It is however striking that the 3-manifold with the string as a boundary,
introduced in ref. [\WessZuminoWitten], from being a mathematical
construction, now obtains a concrete reality as the membrane world-volume.
One \ae sthetic feature is that instead of introducing a separate
E${}_8$ for each boundary string, one naturally gets a single E${}_8$
on the union of the boundaries.

There are a number of questions that remain unanswered. One is the 
microscopic M-theory calculation verifying the exact form of the
Lorentz anomalies that was left out in ref. [\Horava] (although 
specific predictions were made). Another one is the question to which
extent this modified supermembrane theory can be seen as a
microscopic origin of the boundary \SYM. It is conceivable that
the boundary string theory, due to its bulk interaction becomes
``critical'', and that its massless spectrum contains exactly the
10-dimensional \SYM\ fields. The analogous question could be raised
for supermembranes ending on M-theory 5-branes [\Fivebranes], where
the massless selfdual tensor multiplet is believed to be related
to a 6-dimensional string theory without gravitational coupling.
There seems to be a deep connection between M-theory and selfduality
in various dimensions.

\vfill\eject
\refout
\end